\documentclass[aps,showpacs]{revtex4}
\usepackage{graphicx}
\begin{document}

\title{The centrality dependence of transverse energy and charged particle
multiplicity at RHIC: Statistical model analysis}
\author{Dariusz Prorok}
\email{prorok@ift.uni.wroc.pl} \affiliation{Institute of
Theoretical Physics, University of Wroc{\l}aw,\\ Pl.Maksa Borna 9,
50-204 Wroc{\l}aw, Poland}
\date{August 5, 2005}

\begin{abstract}
The transverse energy and charged particle multiplicity at
midrapidity and their ratio are evaluated in a statistical model
with the longitudinal and transverse flows for different
centrality bins at RHIC at $\sqrt{s_{NN}}=130$ and 200 GeV. Full
description of decays of hadron resonances is applied in these
estimations. The predictions of the model at the freeze-out
parameters, which were determined independently from measured
particle yields and $p_{T}$ spectra, agree qualitatively well with
the experimental data. The observed overestimation of the ratio
can be explained for more central collisions by the redefinition
of $dN_{ch}/d\eta\vert_{mid}$.
\end{abstract}

 \maketitle

\section {Introduction}

In the previous paper \cite{Prorok:2004af} the extensive analysis
of two measured global variables, transverse energy
($dE_{T}/d\eta\vert_{mid}$) and charged particle multiplicity
($dN_{ch}/d\eta\vert_{mid}$) densities at midrapidity, was
delivered. The analysis was done in the framework of the single
freeze-out statistical model
\cite{Florkowski:2001fp,Broniowski:2001we,Broniowski:2001uk} for
the most central collision cases of AGS, SPS and RHIC. Now the
same method will be applied in the examination of the centrality
dependence of the above-mentioned variables and their ratio. The
main idea of this method is as follows. Thermal and geometric
parameters of the model are determined from fits to the particle
yield ratios and $p_{T}$ spectra, respectively. Then, with the use
of these parameters both densities, $dE_{T}/d\eta$ and
$dN_{ch}/d\eta$, can be estimated numerically and compared with
the data. The first part of this two-step prescription has been
already done for the existing midrapidity data for different
centrality bins at RHIC at $\sqrt{s_{NN}}=130$ and 200~GeV
\cite{Broniowski:2002nf,Baran:2003nm}. In the present paper the
second step will be performed, namely the estimations of
$dE_{T}/d\eta\vert_{mid}$ and $dN_{ch}/d\eta\vert_{mid}$ for these
centralities and comparison with the data reported in
Refs.~\cite{Adcox:2000sp,Adcox:2001ry,Adams:2004cb,Adler:2004zn}.
The main reason for doing it is that the transverse energy and
charged particle multiplicity measurements are independent of
hadron spectroscopy (in particular, no particle identification is
necessary), therefore they could be used as an additional test of
the self-consistency of a statistical model. There is also an
additional pragmatic reason: predictions of variety of theoretical
models were confronted with the data in \cite{Adler:2004zn}, but
none of these models was a statistical model.

The experimentally measured transverse energy is defined as

\begin{equation}
E_{T} = \sum_{i = 1}^{L} \hat{E}_{i} \cdot \sin{\theta_{i}} \;,
\label{Etdef}
\end{equation}
%%%%%%%%%%%%%%%%%%%%%%%%%%%%Eq.1

\noindent where $\theta_{i}$ is the polar angle, $\hat{E}_{i}$
denotes $E_{i}-m_{N}$ ($m_{N}$ means the nucleon mass) for
baryons, $E_{i}+m_{N}$ for antibaryons and the total energy
$E_{i}$ for all other particles, and the sum is taken over all $L$
emitted particles \cite{Adler:2004zn}.

As a statistical model the single freeze-out model is applied (for
details see \cite{Broniowski:2002nf}). The model succeeded in the
accurate description of ratios and $p_{T}$ spectra of particles
measured at RHIC
\cite{Florkowski:2001fp,Broniowski:2001we,Broniowski:2001uk}. The
main postulate of the model is the simultaneous occurrence of
chemical and thermal freeze-outs, which means that the possible
elastic interactions after the chemical freeze-out are neglected.
The conditions for the freeze-out are expressed by values of two
independent thermal parameters: $T$ and $\mu_{B}$. The strangeness
chemical potential $\mu_{S}$ is determined from the requirement
that the overall strangeness equals zero.

The second basic feature of the model is the complete treatment of
resonance decays. This means that the final distribution of a
given particle consists not only of the thermal part but also of
contributions from all possible decays and cascades.

\section { Foundations of the single-freeze-out model }
\label{Foundat}

The main assumptions of the model are as follows. A noninteracting
gas of stable hadrons and resonances at chemical and thermal
equilibrium appears at the latter stages of a heavy-ion collision.
The gas cools and expands, and after reaching the freeze-out point
it ceases. The chemical and thermal freeze-outs take place
simultaneously. All confirmed resonances up to a mass of $2$ GeV
from the Particle Data Tables \cite{Hagiwara:fs}, together with
stable hadrons, are constituents of the gas. The freeze-out
hypersurface is defined by the equation

\begin{equation}
\tau = \sqrt{t^{2}-r_{x}^{2}-r_{y}^{2}-r_{z}^{2}}= const \;.
\label{Hypsur}
\end{equation}
%%%%%%%%%%%%%%%%%%%%%%%%%%%%Eq.2

\noindent The four-velocity of an element of the freeze-out
hypersurface is proportional to its coordinate

\begin{equation}
u^{\mu}={ {x^{\mu}} \over \tau}= {t \over \tau}\; \left(1,{
{r_{x}} \over t},{{r_{y}} \over t},{{r_{z}} \over t}\right) \;.
\label{Velochyp}
\end{equation}
%%%%%%%%%%%%%%%%%%%%%%%%%%%%Eq.3

\noindent The following parameterization of the hypersurface is
chosen:

\begin{equation}
t= \tau \cosh{\alpha_{\parallel}}\cosh{\alpha_{\perp}},\;\;\;
r_{x}=  \tau \sinh{\alpha_{\perp}}\cos{\phi},\;\;\; r_{y}=  \tau
\sinh{\alpha_{\perp}}\sin{\phi},\;\;\;r_{z}=\tau
\sinh{\alpha_{\parallel}}\cosh{\alpha_{\perp}}, \label{Parahyp}
\end{equation}
%%%%%%%%%%%%%%%%%%%%%%%%%%%%eq.4

\noindent where $\alpha_{\parallel}$ is the rapidity of the
element, $\alpha_{\parallel}= \tanh^{-1}(r_{z}/t)$, and
$\alpha_{\perp}$ controls the transverse radius:

\begin{equation}
r= \sqrt{r_{x}^{2}+r_{y}^{2}}= \tau \sinh{\alpha_{\perp}}.
\label{Transsiz}
\end{equation}

\noindent The transverse size is restricted by the condition $r <
\rho_{max}$. This means that two new parameters of the model have
been introduced, \emph{i.e.} $\tau$ and $\rho_{max}$, which are
connected with the geometry of the freeze-out hypersurface.

The invariant distribution of the measured particles of species
$i$ has the form \cite{Broniowski:2001we,Broniowski:2001uk}

\begin{equation}
{ {dN_{i}} \over {d^{2}p_{T}\;dy} }=\int
p^{\mu}d\sigma_{\mu}\;f_{i}(p \cdot u) \;, \label{Cooper}
\end{equation}
%%%%%%%%%%%%%%%%%%%%%%%%%%%%Eq.9

\noindent where $d\sigma_{\mu}$ is the normal vector on a
freeze-out hypersurface, $p \cdot u = p^{\mu}u_{\mu}$ , $u_{\mu}$
is the four-velocity of a fluid element and $f_{i}$ is the final
momentum distribution of the particle in question. The final
distribution means here that $f_{i}$ is the sum of primordial and
simple and sequential decay contributions to the particle
distribution. The primordial part of $f_{i}$ is given by a
Bose-Einstein or a Fermi-Dirac distribution at the freeze-out. A
decay contribution is a one-dimensional or multidimensional
integral of the momentum distribution of a decaying resonance (the
exact formulae are obtained from the elementary kinematics of a
many-body decay or the superposition of such decays, for details
see \cite{Broniowski:2002nf} and the Appendix in
\cite{Prorok:2004af}). The resonance is a constituent of the
hadron gas and its distribution is also given by the Bose-Einstein
(Fermi-Dirac) distribution function. Therefore the final
distribution $f_{i}$ depends explicitly on $T$ and $\mu_{B}$.

With the use of eqs.~(\ref{Velochyp}) and (\ref{Parahyp}), the
invariant distribution (\ref{Cooper}) takes the following form:

\begin{equation}
{ {dN_{i}} \over {d^{2}p_{T}\;dy} }= \tau^{3}\;
\int\limits_{-\infty}^{+\infty}
d\alpha_{\parallel}\;\int\limits_{0}^{\rho_{max}/\tau}\;\sinh{\alpha_{\perp}}
d(\sinh{\alpha_{\perp}})\; \int\limits_{0}^{2\pi} d\xi\;p \cdot u
\; f_{i}(p \cdot u) \;, \label{Cooper2}
\end{equation}
%%%%%%%%%%%%%%%%%%%%%%%%%%%%eq.10

\noindent where

\begin{equation}
p \cdot u = m_{T}\cosh{\alpha_{\parallel}}\cosh{\alpha_{\perp}}-
p_{T}\cos{\xi}\sinh{\alpha_{\perp}}\;. \label{Peu}
\end{equation}

\noindent  Note that the above distribution is explicitly boost
invariant.

The pseudorapidity density of particle species $i$ is given by

\begin{equation}
{ {dN_{i}} \over {d\eta} } = \int d^{2}p_{T}\; {{dy} \over {d\eta}
} \; { {dN_{i}} \over {d^{2}p_{T}\;dy} }= \int d^{2}p_{T}\; {p
\over {E_{i}} } \; { {dN_{i}} \over {d^{2}p_{T}\;dy} }\;.
\label{Partdens}
\end{equation}
%%%%%%%%%%%%%%%%%%%%%%%%%%%%Eq.13

\noindent Analogously, the transverse energy pseudorapidity
density for the same species can be written as

\begin{equation}
{ {dE_{T,i}} \over {d\eta} } = \int d^{2}p_{T}\; \hat{E}_{i} \cdot
{{p_{T}} \over p} \; {{dy} \over {d\eta} }\; { {dN_{i}} \over
{d^{2}p_{T}\;dy} }= \int d^{2}p_{T}\;{p_{T}} \; { {\hat{E}_{i}}
\over {E_{i}} }\; { {dN_{i}} \over {d^{2}p_{T}\;dy} }\;.
\label{Etraden}
\end{equation}
%%%%%%%%%%%%%%%%%%%%%%%%%%%%Eq.14

\noindent For the quantities at midrapidity one has (in the
c.m.s., which is the RHIC frame of reference)

\begin{equation}
{ {dN_{i}} \over {d\eta} }\;\Big\vert_{mid}= \int d^{2}p_{T}\; {
{p_{T}} \over {m_{T}} } \;{ {dN_{i}} \over {d^{2}p_{T}\;dy} }\; ,
\label{Partdenmid}
\end{equation}
%%%%%%%%%%%%%%%%%%%%%%%%%%%%Eq.15

\begin{equation}
{ {dE_{T,i}} \over {d\eta} }\;\Big\vert_{mid} = \cases{ \int
d^{2}p_{T}\;{p_{T}} \; { {m_{T}-m_{N}} \over {m_{T}} }\; {
{dN_{i}} \over {d^{2}p_{T}\;dy} }\;, i=baryon
 \cr \cr \int
d^{2}p_{T}\;{p_{T}}\; { {m_{T}+m_{N}} \over {m_{T}} } \;{ {dN_{i}}
\over {d^{2}p_{T}\;dy} }\;, i=antibaryon
 \cr \cr \int d^{2}p_{T}\;{p_{T}} \;{ {dN_{i}} \over
{d^{2}p_{T}\;dy} }, i=others  \;.} \label{Etdenmid}
\end{equation}
%%%%%%%%%%%%%%%%%%%%%%%%%%%%Eq.16

\noindent Note that for the older data from RHIC at
$\sqrt{s_{NN}}=130$ GeV \cite{Adcox:2001ry} the \emph{antibaryon}
case is not distinguished and the corresponding expression for the
transverse energy density is the same as for all other particles.

The overall charged particle and transverse energy densities can
be expressed as

\begin{equation}
{ {dN_{ch}} \over {d\eta} }\;\Big\vert_{mid}= \sum_{i \in B} {
{dN_{i}} \over {d\eta} }\;\Big\vert_{mid}\;, \label{Nchall}
\end{equation}
%%%%%%%%%%%%%%%%%%%%%%%%%%%%Eq.18

\begin{equation}
{ {dE_{T}} \over {d\eta} }\;\Big\vert_{mid}= \sum_{i \in A} {
{dE_{T,i}} \over {d\eta} }\;\Big\vert_{mid} \;, \label{Etall}
\end{equation}
%%%%%%%%%%%%%%%%%%%%%%%%%%%%Eq.19

\noindent where $A$ and $B$ ($B \subset A$) denote sets of species
of finally detected particles, namely the set of charged particles
$B$ comprises $\pi^{+},\; \pi^{-},\; K^{+},\; K^{-},\; p$ and
$\bar{p}$, whereas $A$ also includes photons, $K_{L}^{0},\; n$ and
$\bar{n}\;$ \cite{Adcox:2001ry}.

\section {Summary of the fit procedure}
\label{Fitting}

Analyses of the particle ratios and $p_{T}$ spectra at various
centralities in the framework of the single freeze-out model were
done for RHIC in \cite{Broniowski:2002nf,Baran:2003nm}. Here is
the brief summary of this approach. It proceeds in two steps.
First, thermal parameters $T$ and $\mu_{B}$ are fitted with the
use of the experimental ratios of hadron multiplicities at
midrapidity. After then two next parameters, $\tau$ and
$\rho_{max}$, are determined from the simultaneous fit to the
transverse-momentum spectra of $\pi^{\pm}$, $K^{\pm}$, $p$ and
$\bar{p}$. The fits are performed with the help of the $\chi^{2}$
method. For the $k$th measured quantity $R_{k}^{exp}$ and its
theoretical equivalent $R_{k}^{th}(\alpha_{1},...,\alpha_{l})$,
which depends on $l$ parameters $\alpha_{1},...,\alpha_{l}$, the
$\chi^{2}$ function is defined as

\begin{equation}
\chi^{2}(\alpha_{1},...,\alpha_{l}) = \sum_{k=1}^{n} {
(R_{k}^{exp}-R_{k}^{th}(\alpha_{1},...,\alpha_{l}))^{2} \over
\sigma_{k}^{2}} \;, \label{Chidef}
\end{equation}

\noindent where $\sigma_{k}$ is the error of the $k$th measurement
and $n$ is the total number of data points. The fitted values of
parameters mean the values at which $\chi^{2}$ has a minimum.

To determine $T$ and $\mu_{B}$ the $k$th measured ratio of hadron
multiplicities at midrapidity is put into eq.~(\ref{Chidef}) as
the measured quantity:

\begin{equation}
R_{k}^{exp} = { dN_{i}^{exp}/dy  \over dN_{j}^{exp}/dy
}\Big\vert_{mid} \;. \label{Ratmult}
\end{equation}

\noindent In the case of a boost-invariant model (as here), the
theoretical equivalent $R_{k}^{th}(T, \mu_{B})$ is given by

\begin{equation}
R_{k}^{th}(T, \mu_{B}) = { dN_{i}^{th}/dy  \over dN_{j}^{th}/dy
}\Big\vert_{mid}  = { N_{i}^{th} \over N_{j}^{th} } = { n_{i}(T,
\mu_{B}) \over n_{j}(T, \mu_{B}) } \;, \label{Rattheor}
\end{equation}

\noindent where $n_{i}(T, \mu_{B})$ is the final density of
particle species $i$ calculated for a static gas. The last
equality follows from the assumption that the temperature and
chemical potentials are constant on the freeze-out hypersurface.
Then the volume of the hypersurface factorizes when the invariant
distribution (\ref{Cooper}) is integrated formally over all
momenta to obtain the total multiplicity $N_{i}^{th}$ (for details
see \cite{Broniowski:2002nf}). The final density means here that
it collects all decay contributions. Thus the final density of
particle species $i$ reads

\begin{equation}
n_{i}(T, \mu_{B}) = n_{i}^{primordial}(T, \mu_{B}) + \sum_{a}
\varrho(i,a)\; n_{a}^{primordial}(T, \mu_{B}) \;, \label{nchj}
\end{equation}
%%%%%%%%%%%%%%%%%%%%%%%%%%%%eq.5

\noindent where $n_{a}^{primordial}(T, \mu_{B})$ is the thermal
density of the $a$th particle species at the freeze-out,
$\varrho(i,a)$ is the final number of particles of species $i$
which can be received from all possible simple or sequential
decays of particle $a$ and the sum is over all species of
resonances included in the hadron gas. The values of $T$ and
$\mu_{B}$ were fixed for both RHIC energies (\textit{i.e.}
$\sqrt{s_{NN}}=130$ and 200 GeV) with the use of data for the most
central collisions \cite{Broniowski:2002nf,Baran:2003nm}. In the
further considerations it is assumed that these values are
independent of the centrality. This is reasonable since the very
weak centrality dependence of the particle ratios has been
observed so far. Recent analyses done in
\cite{Cleymans:2004pp,Rafelski:2004dp} have just confirmed the
above-mentioned assumption.

The second step of this approach is to determine values of $\tau$
and $\rho_{max}$. Now the $k$th measured quantity $R_{k}^{exp}$ is
the value of $dN_{i}^{exp}/(d^{2}p_{T}dy)\vert_{mid}$ for measured
particle species $i$ and its transverse momentum $p_{T}$, whereas
the theoretical equivalent $R_{k}^{th}(\tau, \rho_{max})$ is given
by the formula (\ref{Cooper2}). Note that again the $\chi^{2}$
function depends on two free parameters, now $\tau$ and
$\rho_{max}$, since the values of $T$ and $\mu_{B}$, which were
determined early, have been put into eq.~(\ref{Cooper2}).

The fitted values of parameters of the model are taken from
\cite{Broniowski:2002nf,Baran:2003nm} and are gathered in
table~\ref{Table1}. Since not all bins reported in
\cite{Adcox:2001mf,Chujo:2002bi,Barannikova:2002qw} were examined,
the lacking values of the geometric parameters have been obtained
from the linear approximation between the nearest up and down
neighbors. This is justified because the geometric parameters,
when plotted as a function of centrality, show roughly linear
dependence \cite{Baran:2003nm}. Also in table~\ref{Table1} the
corresponding number of participants ($N_{part}$) is given for
each bin. If the division into centrality classes is different for
the identified charged hadron measurements and the $dE_{T}/d\eta$
and $dN_{ch}/d\eta$ measurements, values of $N_{part}$ are taken
from reports on the former. Therefore  values of $N_{part}$ from
\cite{Adcox:2001mf,Adler:2003cb} are listed for PHENIX, whereas
values of $N_{part}$ for STAR are taken from its transverse energy
measurement analysis \cite{Adams:2004cb} (in
\cite{Chujo:2002bi,Barannikova:2002qw} numbers of participants are
not given).

%%%%%%%%%%%%%%
\begin{table}
\caption{\label{Table1} Values of thermal and geometric parameters
of the model for various centrality bins taken from
\protect\cite{Broniowski:2002nf,Baran:2003nm}. Marked are these
bins, for which values of geometric parameters have been obtained
by the author from linear approximation between the nearest
neighbours (see the text for the explanation). }
\begin{ruledtabular}
\begin{tabular}{ccccc} \hline Collision case & Centrality [\%] &
$N_{part}$ & $\tau$ [fm] & $\rho_{max}$ [fm]
\\
\hline PHENIX at $\sqrt{s_{NN}}=130$ GeV: & 0-5 & 348 & 8.20 &
6.90
\\
$T = 165$ MeV, $\mu_{B} = 41$ MeV & 5-15$^{*}$ & 271 & 7.49 & 6.30
\\
 & 15-30 & 180 & 6.30 & 5.30
\\
 & 30-60$^{*}$ & 79 & 4.62 & 3.91
\\
 & 60-92 & 14 & 2.30 & 2.0
\\
\hline PHENIX at $\sqrt{s_{NN}}=200$ GeV: & 0-5 & 351.4 & 7.86 &
7.15
\\
$T = 165.6$ MeV, $\mu_{B} = 28.5$ MeV & 5-10$^{*}$ & 299.0 & 7.48
& 6.81
\\
 & 10-15$^{*}$ & 253.9 & 7.10 & 6.47
\\
 & 15-20$^{*}$ & 215.3 & 6.71 & 6.13
\\
 & 20-30 & 166.6 & 6.14 & 5.62
\\
 & 30-40 & 114.2 & 5.73 & 4.95
\\
 & 40-50 & 74.4 & 4.75 & 3.96
\\
 & 50-60 & 45.5 & 3.91 & 3.12
\\
 & 60-70 & 25.7 & 3.67 & 2.67
\\
 & 70-80 & 13.4 & 3.09 & 2.02
\\
\hline STAR at $\sqrt{s_{NN}}=200$ GeV: & 0-5 & 352 & 9.74 & 7.74
\\
$T = 165.6$ MeV, $\mu_{B} = 28.5$ MeV & 5-10 & 299 & 8.69 & 7.18
\\
 & 10-20 & 234 & 8.12 & 6.44
\\
 & 20-30 & 166 & 7.24 & 5.57
\\
 & 30-40 & 115 & 7.07 & 4.63
\\
 & 40-50 & 76 & 6.38 & 3.91
\\
 & 50-60 & 47 & 6.19 & 3.25
\\
 & 60-70$^{*}$ & 27 & 5.70 & 2.51
\\
 & 70-80 & 14 & 5.21 & 1.76
\\
\hline
\end{tabular}
\end{ruledtabular}
\end{table}
%%%%%%%%%%%%%%%%%%%%%%%%%%%%%%%%%

\section {Results}
\label{Finl}

\begin{figure}
\includegraphics[width=0.55\textwidth]{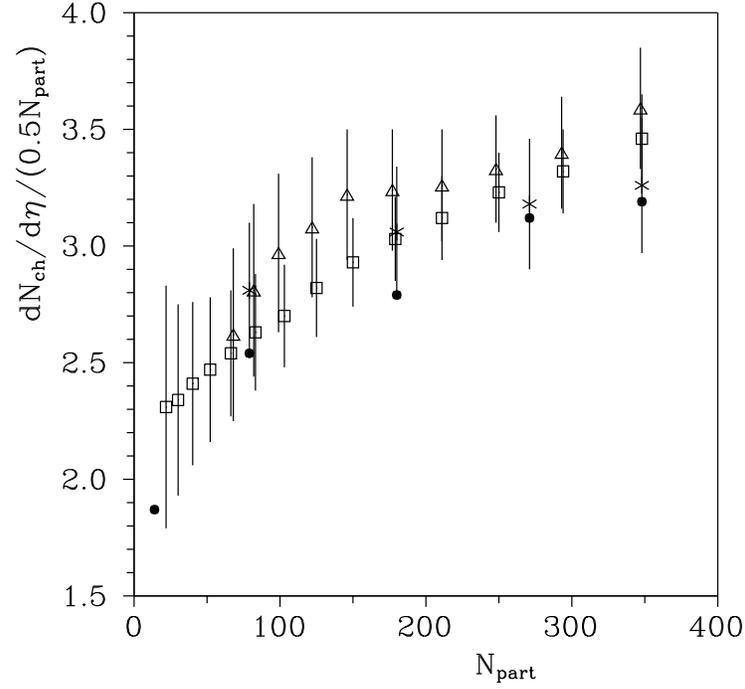}
\caption{\label{Fig.1} $dN_{ch}/d\eta$ per pair of participants
versus $N_{part}$ for RHIC at $\sqrt{s_{NN}}=130$ GeV. Dots denote
model evaluations, squares the newest PHENIX data
\protect\cite{Adler:2004zn}, triangles the earlier reported PHENIX
data \protect\cite{Adcox:2000sp} and crosses are the recalculated
PHENIX data from summing up the integrated charged hadron yields
delivered in \protect\cite{Adcox:2001mf}. }
\end{figure}

\begin{figure}
\includegraphics[width=0.55\textwidth]{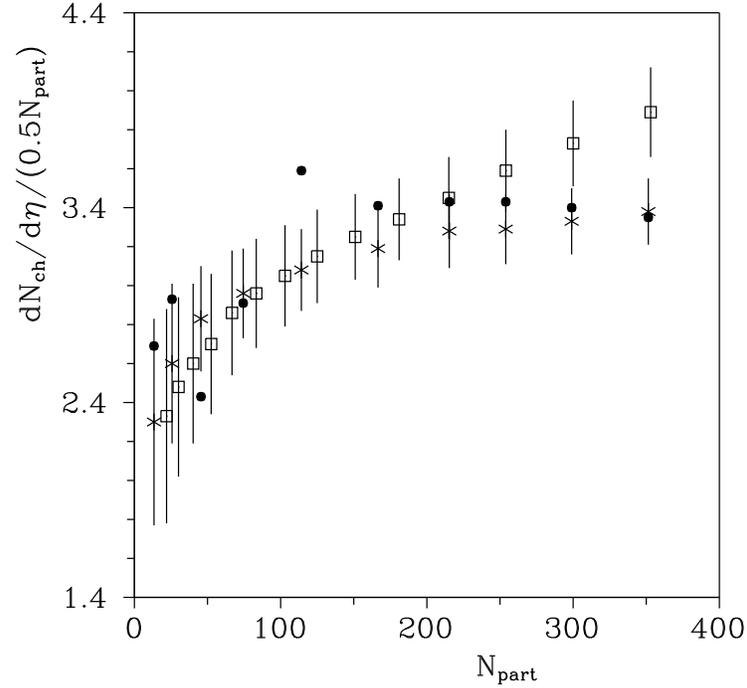}
\caption{\label{Fig.2} $dN_{ch}/d\eta$ per pair of participants
versus $N_{part}$ for RHIC at $\sqrt{s_{NN}}=200$ GeV. Dots denote
model evaluations, squares PHENIX data \protect\cite{Adler:2004zn}
and crosses are the recalculated PHENIX data from summing up the
integrated charged hadron yields delivered in
\protect\cite{Adler:2003cb}. }
\end{figure}

The results of numerical estimations of
$dN_{ch}/d\eta\vert_{mid}$, eq.~(\ref{Nchall}), divided by the
number of participant pairs for PHENIX centrality bins tabulated
in table~\ref{Table1} are presented in figs.\,\ref{Fig.1} and
\ref{Fig.2} for $\sqrt{s_{NN}}=130$ and 200 GeV respectively.
Additionally to the straightforward PHENIX measurements of the
total charged particle multiplicity, also the data from the
summing up of the integrated charged hadron yields are depicted in
these figures (more precisely, since the integrated charged hadron
yields are given as rapidity densities, the transformation to
pseudo-rapidity should be done, which means the division by a
factor 1.2 here, see \cite{Bazilevsky:2002fz}). This is because
fits of the parameters of the model should be done to the same
$p_{T}$ spectra which are to be integrated to delivered the
charged hadron yields. For PHENIX at $\sqrt{s_{NN}}=200$ GeV this
is not true since fits were done to the preliminary data
\cite{Chujo:2002bi}, but the integrated charged hadron yields were
delivered in the final report \cite{Adler:2003cb}. And as one can
see from a comparison between figures in \cite{Chujo:2002bi} and
\cite{Adler:2003cb}, points from the former are slightly above
corresponding points from the latter. To estimate the scale of
this difference, the sum of integrated charged hadron yields at
$N_{part}=114.2$ (the point of the biggest discrepancy between the
model evaluation and the experimental data, see fig.\,\ref{Fig.2})
has been obtained again from digitizing the data depicted in
preliminary plots in \cite{Chujo:2002bi}. This gives 3.31 of
charged particles per participant pair, which is exactly $10\%$
above the value obtained from the data given in
\cite{Adler:2003cb}. And the model evaluation is only $8.5\%$
greater than this number. Also in \cite{Adler:2003cb} the feeding
of $p(\bar{p})$ from $\Lambda(\bar{\Lambda})$ decays is excluded,
contrary to \cite{Chujo:2002bi}. Therefore, the model estimates
should overestimate the corresponding recalculated experimental
values for PHENIX at $\sqrt{s_{NN}}=200$ GeV. To diminish this
effect, integrated $p$ and $\bar{p}$ yields delivered in
\cite{Adler:2003cb}, were corrected to include back the feeding.
The correction was done by the division by a factor 0.65, which is
rough average of a $p_{T}$ dependent multiplier used by the PHENIX
Collaboration (see fig.4 and eq.(5) in \cite{Adler:2003cb}). Since
all just stated reasons for the "systematic" discrepancies affect
the absolute values of $dN_{ch}/d\eta\vert_{mid}$, they reveal
themselves in the normalization factor $\tau^{3}$ in the model (
see eq.~(\ref{Cooper2})). For PHENIX at $\sqrt{s_{NN}}=130$ GeV
the same spectra were used to fit the model parameters and to
obtain the integrated yields \cite{Adcox:2001mf}, so the
predictions should agree with the recalculated data in principle.
As one can notice from figs.\,\ref{Fig.1} and \ref{Fig.2}, the
above comments concerning both PHENIX measurements are true. The
greatest overestimation is in the most peripheral bin and in the
mid-centrality region of the measurements at $\sqrt{s_{NN}}=200$
GeV. But even at the worst point the model prediction is $17\%$
over the corresponding recalculated experimental value. It is not
so bad because the absolute value of theoretical $dN_{ch}/d\eta$
is determined by the factor $\tau^{3}$. This means that the
$5.3\%$ uncertainty in $\tau$ is enough to cause the $17\%$
uncertainty in $dN_{ch}/d\eta$. The second reason for the
quantitative disagreement is that transverse momentum spectra are
measured in \emph{limited ranges}, so very important low-$p_{T}$
regions are not covered by the data. To obtain integrated yields
some extrapolations below and above the measured ranges are used.
In fact these extrapolations are only analytical fits, but
contributions from regions covered by them account for about
$25-40\%$ of the integrated yields \cite{Adcox:2001mf}. It might
turned out that these extrapolations differ from the thermal
distributions supplemented by the distributions of products of
decays.

%%%%%%%%%%%%%%%%%%%%

Generally, the agreement of the model predictions with the data is
much better for RHIC at $\sqrt{s_{NN}}=130$ GeV. For the case of
$\sqrt{s_{NN}}=200$ GeV, only the rough qualitative agreement has
been reached and the reasons have just been explained. It is also
worth to stress once more, that the discrepancy between the
directly measured $dN_{ch}/d\eta$ and $dN_{ch}/d\eta$ expressed as
the sum of the integrated charged hadron yields can be one of
these reasons, especially for RHIC at $\sqrt{s_{NN}}=200$ GeV (see
fig\,\ref{Fig.2}; this effect has already been notified in backup
slides of \cite{Chujo:2002bi}). The discrepancy starts at
mid-centrality and rises with the centrality.

\begin{figure}
\includegraphics[width=0.55\textwidth]{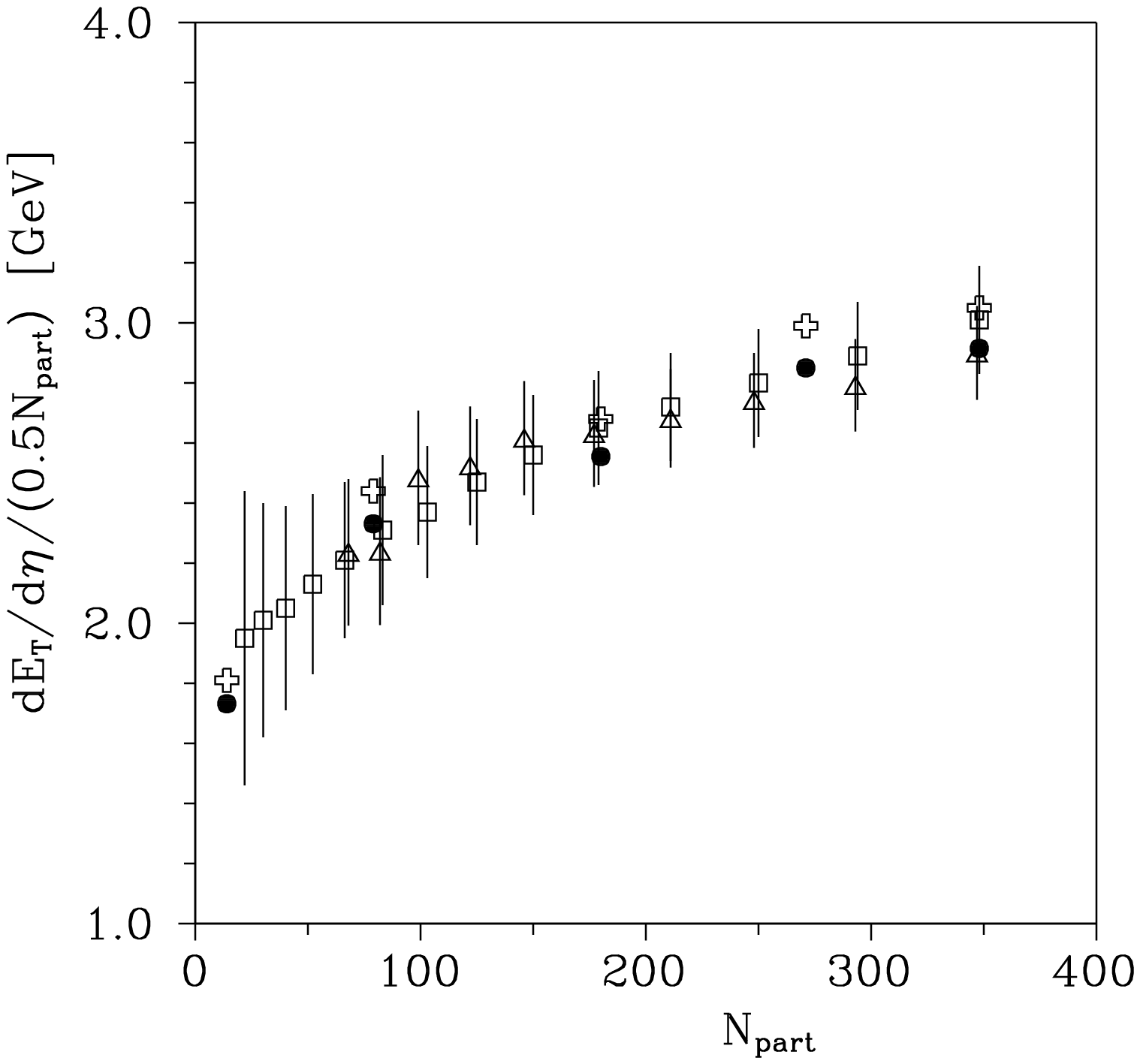}
\caption{\label{Fig.3} $dE_{T}/d\eta$ per pair of participants
versus $N_{part}$ for RHIC at $\sqrt{s_{NN}}=130$ GeV. Dots and
open crosses denote model evaluations, triangles and squares are
PHENIX data \protect\cite{Adcox:2000sp,Adler:2004zn}. Dots and
triangles are for the older definition of $E_{T}$, \emph{i.e.} the
total energy $E_{i}$ is taken also for antibaryons in
eq.~(\ref{Etdef}). }
\end{figure}
%%%%%%%%%%%%%%%%%%%%
\begin{figure}
\includegraphics[width=0.55\textwidth]{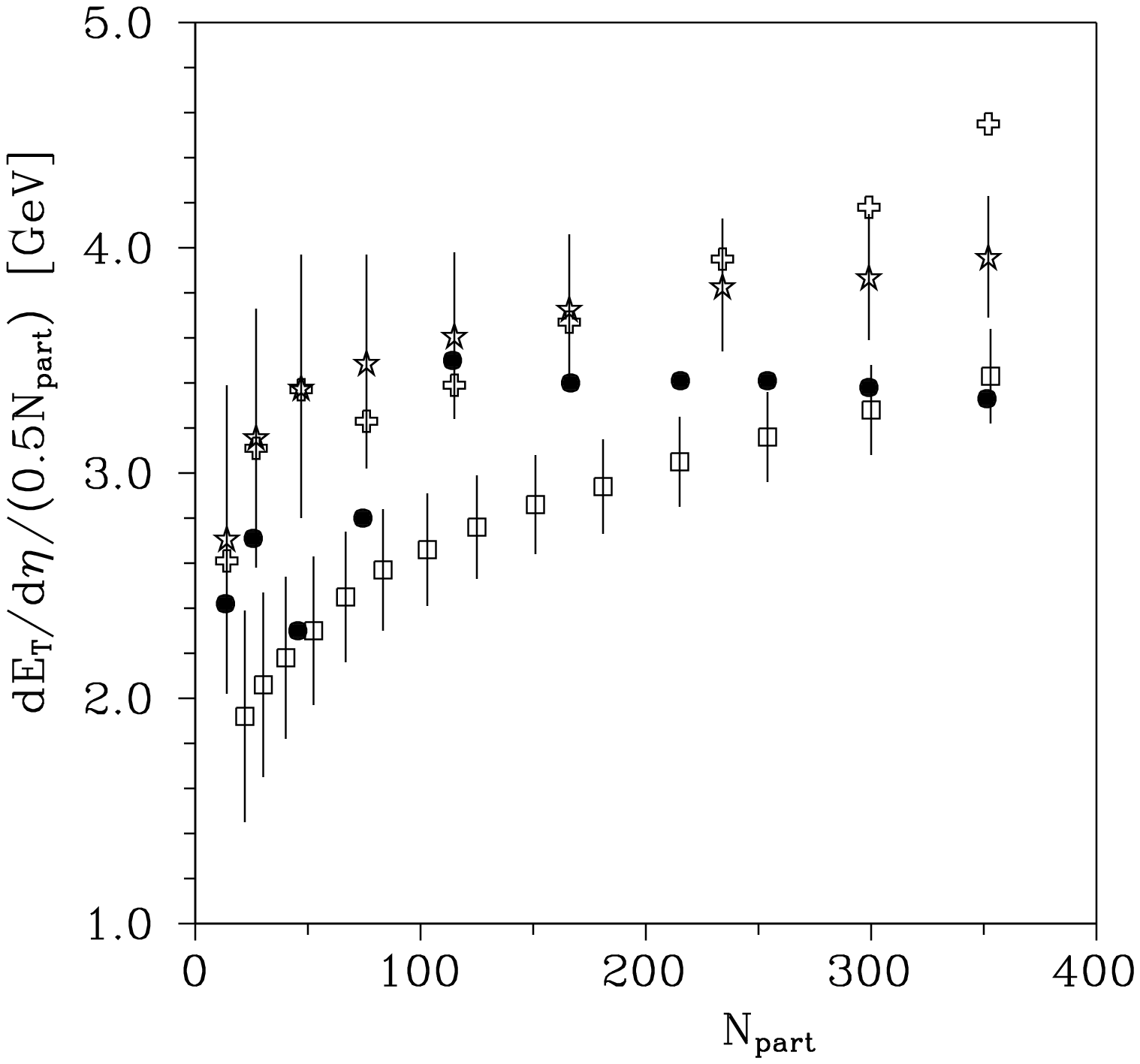}
\caption{\label{Fig.4} $dE_{T}/d\eta$ per pair of participants
versus $N_{part}$ for RHIC at $\sqrt{s_{NN}}=200$ GeV. Dots and
open crosses denote model evaluations, squares and stars are data.
Dots and squares are for PHENIX (data from
\protect\cite{Adler:2004zn}), open crosses and stars for STAR
(original data from \protect\cite{Adams:2004cb} have been rescaled
to $\eta=0$, see the text for the explanation). }
\end{figure}

The estimates of $dE_{T}/d\eta\vert_{mid}$, eq.~(\ref{Etall}),
divided by the number of participant pairs are shown in
figs.\,\ref{Fig.3} and \ref{Fig.4} for $\sqrt{s_{NN}}=130$ and 200
GeV respectively. The quality of the model predictions for
$dE_{T}/d\eta$ measured by PHENIX is the same as for
$dN_{ch}/d\eta$. Again, only the qualitative agreement has been
reached in the case of $\sqrt{s_{NN}}=200$ GeV but the
overestimation is higher and equals $30\%$ at most. The main
source of the overestimation seems to be the same as in the
$dN_{ch}/d\eta$ case, namely the uncertainty of fitting the
parameter $\tau$. The STAR measurements need separate comments.
The STAR data were taken not at midrapidity ($\eta=0$) as in the
PHENIX case but at $\eta=0.5$ on the average \cite{Adams:2004cb}.
But the $p_{T}$ spectra used for fits of the geometric parameters
were measured at midrapidity, also in the STAR case
\cite{Barannikova:2002qw}. Therefore one should expect some
"systematic" discrepancy between predictions and the data on the
whole. To remove this effect the original STAR data
\cite{Adams:2004cb} have been divided by a factor
$\sin{(\theta\vert_{\eta=0.5})}\approx0.887$. As it can be seen
from fig.\,\ref{Fig.4}, the predictions and data agree with each
other within errors besides the most central point. But even
there, the discrepancy does not overcome $15\%$. Of course, the
second source of the quantitative disagreement could be the
uncertainty in $\tau$ (in fact, as it will be seen, it seems to be
the main source of the disagreement in all discussed cases).

Fig.\,\ref{Fig.4} also shows that both experimental and
theoretical values of the transverse energy per participant pair
corresponding to the STAR case are $\sim 30 \%$ greater than the
PHENIX ones. As far as the theoretical estimates are concerned
this is the consequence of the higher values of $p_{T}$
distributions of pions, kaons and antiprotons measured by the STAR
Collaboration with respect to PHENIX measurements (it can be seen
directly from careful comparison of spectra given in
\cite{Barannikova:2002qw} and \cite{Chujo:2002bi}, in fact such a
comparison was done in \cite{Baran:2003nm}, see fig.5 there).
However, it is difficult to judge without doubts what is the
reason for the normalization difference between the experimental
data. Additionally, it is interesting that this difference
decreases with the centrality. One of the reasons could be a
different fiducial aperture: in the PHENIX case measurements were
done for $\mid\eta\mid \leq 0.38$ in pseudo-rapidity and
$\triangle\phi = 44.4^{\circ}$ in azimuth, whereas in the STAR
case the pseudo-rapidity range was $0 < \eta < 1$ and the
azimuthal coverage $\triangle\phi = 60^{\circ}$. In both cases the
correction factor for fiducial acceptance is given by
$1/\triangle\eta \cdot 2\pi/\triangle\phi$, which implicitly
assumes the uniform distribution of the raw $E_{T}$ data in the
covered pseudo-rapidity range and in the azimuthal angle. Of
course, it might not be exactly true and the final difference in
the data could reflect slight deviations from this expected
uniformity. Another point is a rather rough technique used by the
author to rescaled the STAR data to $\eta =0$. The division by a
factor $\sin{(\theta\vert_{\eta=0.5})}\approx0.887$ contributes
almost $13$ percentage points to the mentioned $\sim 30 \%$
increase in the normalization. Anyway, the increase in the
normalization of the transverse-energy measurements observed in
the STAR data with respect to the PHENIX data is consistent with
the similar increase in the normalization of the measured spectra.

\begin{figure}
\includegraphics[width=0.55\textwidth]{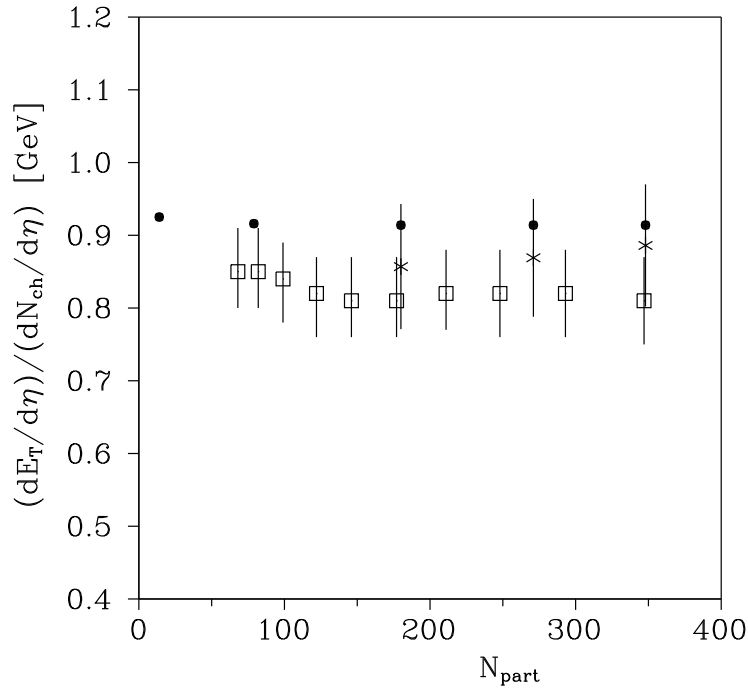}
\caption{\label{Fig.5} $\langle dE_{T}/d\eta\rangle /\langle
dN_{ch}/d\eta\rangle$ versus $N_{part}$ for RHIC at
$\sqrt{s_{NN}}=130$ GeV and for the older definition of $E_{T}$,
\emph{i.e.} the total energy $E_{i}$ is taken also for antibaryons
in eq.~(\ref{Etdef}). Dots denote model evaluations, squares are
the earlier PHENIX data \protect\cite{Adcox:2001ry}. Crosses
denote recalculated PHENIX data points, \emph{i.e.} the sum of
integrated charged hadron yields \cite{Adcox:2001mf} have been
substituted for the denominator in the ratio.}
\end{figure}
%%%%%%%%%%%%%%%%%%%%
\begin{figure}
\includegraphics[width=0.55\textwidth]{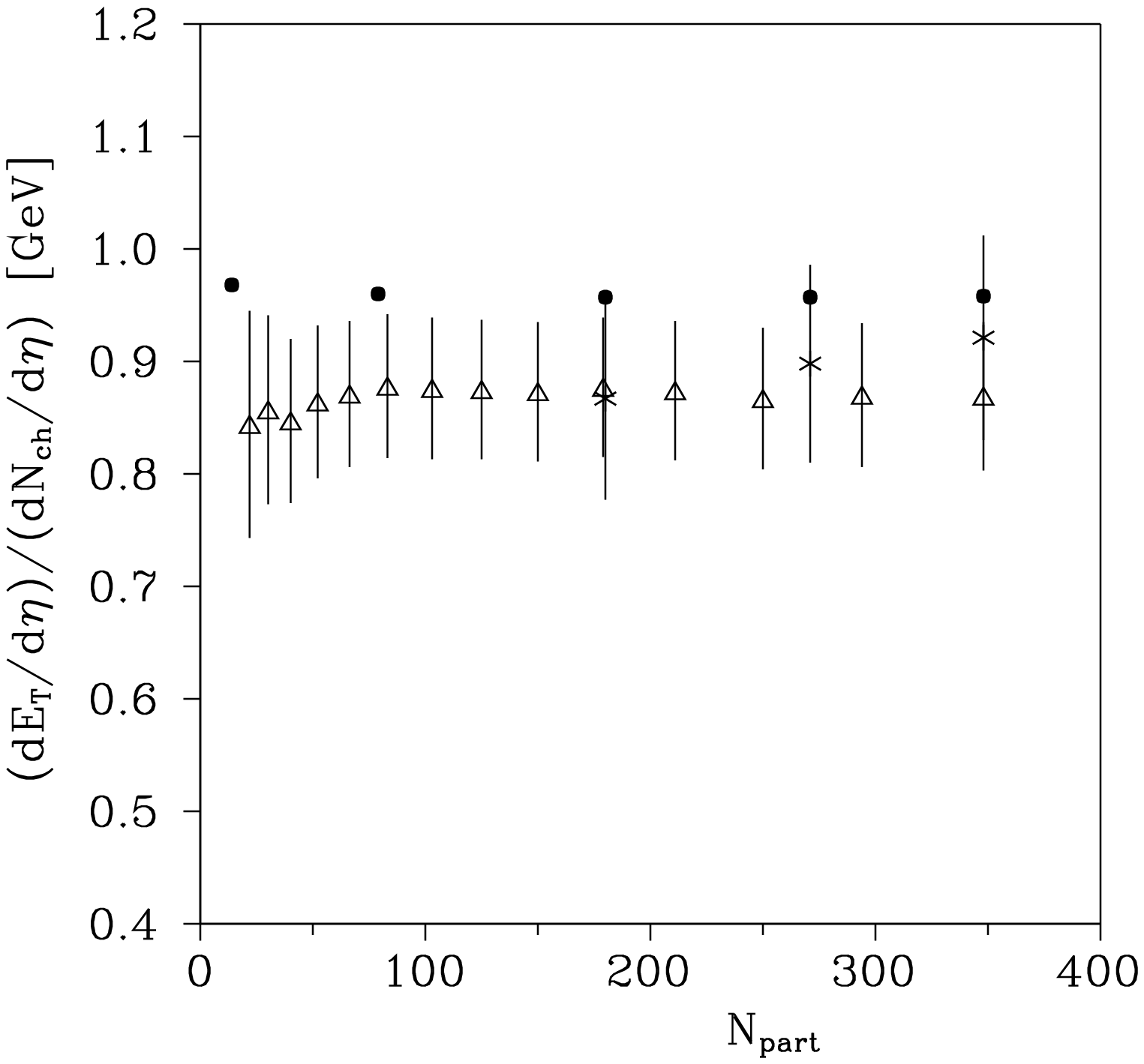}
\caption{\label{Fig.6} $\langle dE_{T}/d\eta\rangle /\langle
dN_{ch}/d\eta\rangle$ versus $N_{part}$ for RHIC at
$\sqrt{s_{NN}}=130$ GeV. Dots denote model evaluations, triangles
are PHENIX data \protect\cite{Adler:2004zn}. Crosses denote
recalculated PHENIX data points, \emph{i.e.} the sum of integrated
charged hadron yields \cite{Adcox:2001mf} have been substituted
for the denominator in the ratio. }
\end{figure}
%%%%%%%%%%%%%%%%%%%%%%%%%%%
\begin{figure}
\includegraphics[width=0.55\textwidth]{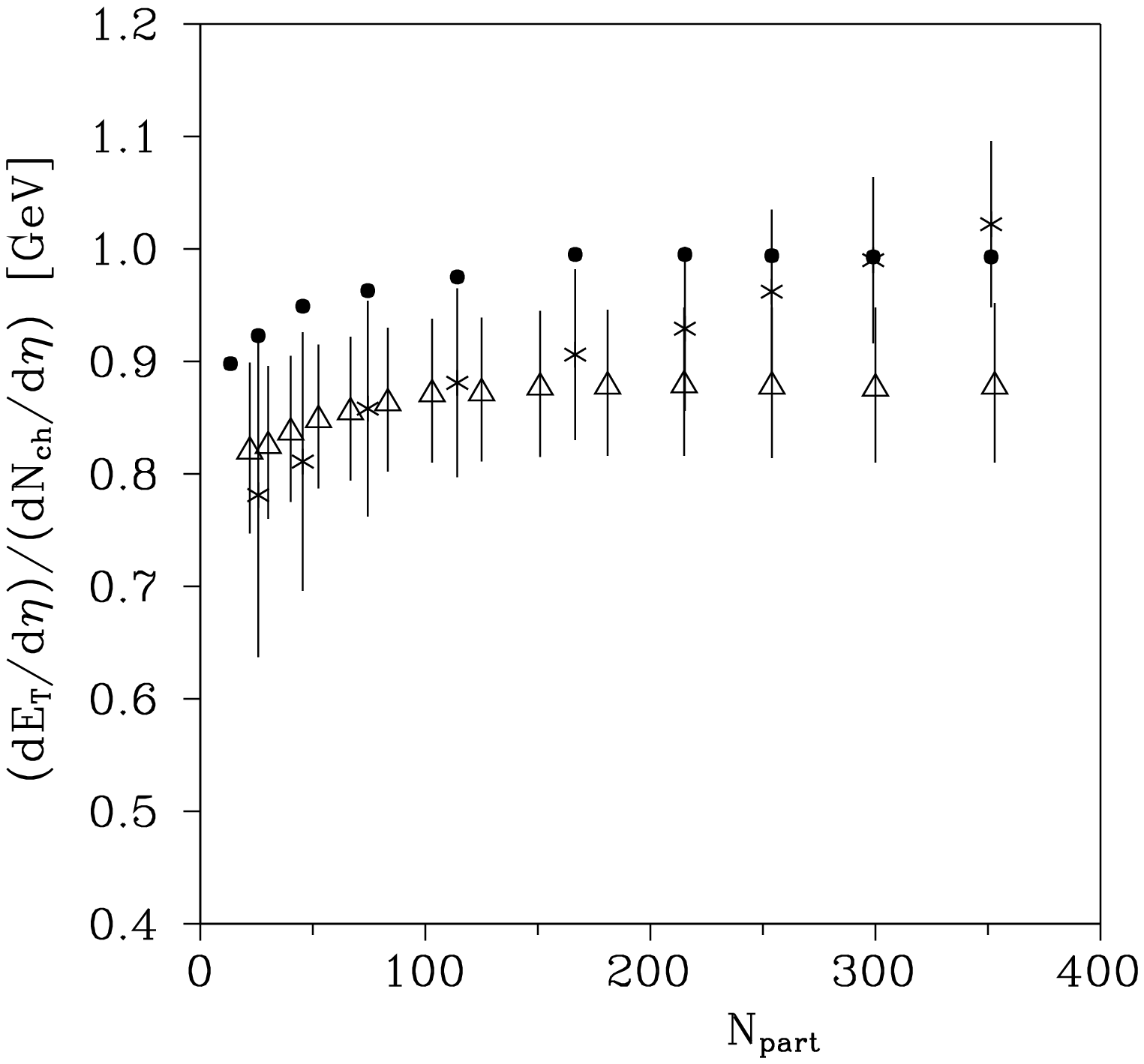}
\caption{\label{Fig.7} $\langle dE_{T}/d\eta\rangle /\langle
dN_{ch}/d\eta\rangle$ versus $N_{part}$ for RHIC at
$\sqrt{s_{NN}}=200$ GeV. Dots denote model evaluations, triangles
are PHENIX data \protect\cite{Adler:2004zn}. Crosses denote
recalculated PHENIX data points, \emph{i.e.} the sum of integrated
charged hadron yields \cite{Adler:2003cb} have been substituted
for the denominator in the ratio. }
\end{figure}
%%%%%%%%%%%%%%%%%
\begin{figure}
\includegraphics[width=0.55\textwidth]{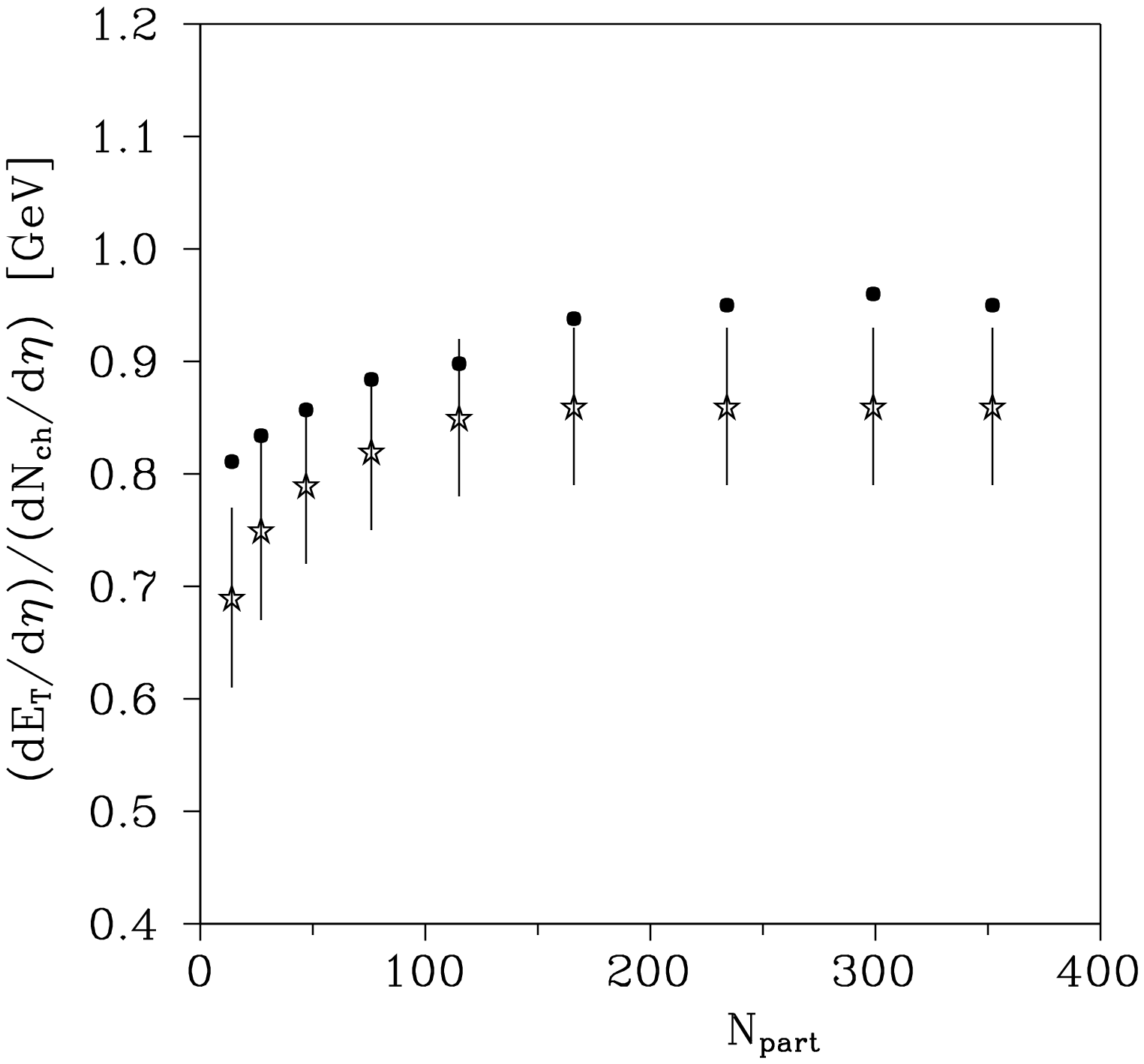}
\caption{\label{Fig.8} $\langle dE_{T}/d\eta\rangle /\langle
dN_{ch}/d\eta\rangle$ versus $N_{part}$ for RHIC at
$\sqrt{s_{NN}}=200$ GeV. Dots denote model evaluations, stars are
STAR data \protect\cite{Adams:2004cb}. }
\end{figure}

Values of the ratio $\langle dE_{T}/d\eta\rangle /\langle
dN_{ch}/d\eta\rangle$ as a function of $N_{part}$ are presented in
figs.\,\ref{Fig.5}-\ref{Fig.8}. As one can see, the position of
model predictions is very regular and exactly resembles the
configuration of the data in each case, the estimates are only
shifted up about $10\%$ as a whole. This indirectly proves that
the earlier discussed disagreement in estimates of
$dN_{ch}/d\eta\vert_{mid}$ and $dE_{T}/d\eta\vert_{mid}$ has its
origin in the uncertainty of fitting the parameter $\tau$ (the
normalization factor $\tau^{3}$ cancels in the ratio). The
observed $10\%$ overestimation of the ratio can be explained, at
least for more central collisions, by the observed discrepancy
between the directly measured $dN_{ch}/d\eta$ and $dN_{ch}/d\eta$
expressed as the sum of the integrated charged hadron yields. If
the original data points are replaced by the recalculated data
such that the denominators are sums of the integrated charged
hadron yields, then much better agreement can be reached for all
but peripheral collisions (see figs.\,\ref{Fig.5}-\ref{Fig.7}).

\section {Conclusions}
\label{Conclud}

The single freeze-out model has been applied to estimate
transverse energy and charged particle multiplicity densities for
different centrality bins of RHIC measurements at
$\sqrt{s_{NN}}=130$ and $200$ GeV. These two variables are
independent observables, which means that they are measured
independently of identified hadron spectroscopy. Since model fits
were done to identified hadron data (particle yield ratios and
$p_{T}$ spectra) and transverse energy and charged particle
multiplicity densities are calculable in the single freeze-out
model, it was very tempting to check whether their estimated
values agree with the data. Generally the answer is yes, at least
on the qualitative level. As it has just turned out, the main
source of the quantitative disagreement is the uncertainty in the
value of the parameter $\tau$. The uncertainty strongly influences
both densities since their theoretical equivalents contain the
normalization factor $\tau^{3}$. This conclusion is confirmed by
the analysis of the transverse energy per charged particle as a
function of the number of participating pairs. The overestimation
of at most $30\%$ obtained for the absolute value of the
transverse energy density decreases to the overall overestimation
of the order of $10\%$ for the ratio $\langle dE_{T}/d\eta\rangle
/\langle dN_{ch}/d\eta\rangle$. On the quantitative level this
means that values of only three parameters of the model are
confirmed in the present analysis, namely $T$, $\mu_{B}$ and the
ratio $\rho_{max}/\tau$. Since the ratio is directly connected
with the maximum transverse-flow parameter (for the derivation see
\cite{Prorok:2004af}),

\begin{equation}
\beta_{\perp}^{max}= { {\rho_{max}/\tau} \over
{\sqrt{1+(\rho_{max}/\tau)^{2}}}}\;, \label{Betmax}
\end{equation}

\noindent this set of parameters is equivalent to $T$, $\mu_{B}$
and $\beta_{\perp}^{max}$, which are more commonly used in other
statistical models describing particle production in heavy-ion
collisions (\emph{e.g.} the blast-wave model
\cite{Schnedermann:1993ws}).

To summarize, the single freeze-out version of a statistical model
fairly well explains the observed centrality dependence of
transverse energy and charged particle multiplicity
pseudo-rapidity densities at midrapidity and their ratio. Also the
dependence on $\sqrt{s_{NN}}$ of the above-mentioned variables is
well recovered \cite{Prorok:2004af}. It should be stressed once
more, that this model very well reproduces the particle ratios and
the transverse-momentum spectra measured at RHIC
\cite{Broniowski:2001we,Broniowski:2001uk,Broniowski:2002nf,Baran:2003nm}.
In fact, the description of the identified hadron data was the
original motivation for the formulation of the model in
\cite{Broniowski:2001we,Broniowski:2001uk}. The results presented
in this paper confirm, in an independent way, the general
conclusion drawn in the above-mentioned references, that the
single freeze-out model is a very useful (and simple) tool for the
estimations of fundamental physical parameters like the
temperature, chemical potentials or the size of the matter created
at the final stages of a heavy-ion collision. This supports the
idea of the appearance of a thermal system during such a
collision.

\begin{acknowledgments}
This work was supported in part by the Polish Committee for
Scientific Research under Contract No. KBN 2 P03B 069 25.
\end{acknowledgments}

\end{document}